\documentclass[pdflatex,sn-mathphys-num]{sn-jnl}

\usepackage{tabularx}
\usepackage{float}
\usepackage{multirow}%
\usepackage{amsthm}%
\usepackage{mathrsfs}%
\usepackage[title]{appendix}%
\usepackage{xcolor}%
\usepackage{textcomp}%
\usepackage{manyfoot}%
\usepackage{algorithm}%
\usepackage{algorithmicx}%
\usepackage{algpseudocode}%
\usepackage{listings}%

\usepackage[utf8]{inputenc}
\usepackage[T1]{fontenc}
\usepackage{amsmath,amssymb,amsfonts} 
\usepackage{graphicx}                
\usepackage{hyperref}                 
\usepackage{booktabs}
\usepackage{setspace}
\usepackage{lineno}

\usepackage{xcolor}
\begin{document}

\title[Article Title]{Optimum foraging area in a three-trophic food chain}

\author*[1,2]{\fnm{Lucas} \sur{Massoni}}\email{lucasmassoni02@gmail.com}

\author[3]{\fnm{Rafael} \sur{Menezes}}\email{r.menezes@ictp-saifr.org}

\author[4]{\fnm{Marcus} \sur{A. M. de Aguiar}}\email{aguiar@ifi.unicamp.br}

\author[2,5]{\fnm{Sabrina} \sur{B.L. Araujo}}\email{araujosbl@ufpr.br}
\equalcont{These authors contributed equally to this work.}

\affil[1]{\orgdiv{Programa de Pós-Graduação em Física}, \orgname{Universidade Federal do Paraná} \orgaddress{
,\city{Curitiba}, 
\state{Paraná}, \country{Brazil}}}

\affil[2]{\orgdiv{Biological Interactions}, \orgname{Universidade Federal do Paraná} \orgaddress{
,\city{Curitiba}, 
\state{Paraná}, \country{Brazil}}}

\affil[3]{\orgdiv{South American Institute for Fundamental Research and Instituto de Física Teórica}, \orgname{Universidade Estadual Paulista}, \orgaddress{
\city{São Paulo}, 
\state{São Paulo}, \country{Brazil}}}

\affil[4]{\orgdiv{Instituto de Física Gleb Wataghin}, \orgname{Universidade Estadual de Campinas}, \orgaddress{
\city{Campinas}, 
\state{São Paulo}, \country{Brazil}}}

\affil[5]{\orgdiv{Departamento de Física}, \orgname{Universidade Federal do Paraná} \orgaddress{
,\city{Curitiba}, 
\state{Paraná}, \country{Brazil}}}

\abstract{
Organisms' foraging strategies are shaped by a trade-off between search area and local capture efficiency. This trade-off leads individuals to adapt their foraging area to an optimal value, impacting population dynamics.  Here we study the effect of multiple foraging areas in a predator-prey model composed of three trophic levels. The interactions between predators and prey occur only within a limited neighborhood of the predators, where adaptation can occur over generations. We assume a trade-off where local predation efficiency is inversely proportional to the foraging area. These dynamics were implemented computationally via cellular automata and analytically via Master Equations with mean-field and pair approximations. Unlike the mean-field approximation, the pair approximation reproduced the dependence of population density on foraging area observed in the simulations. However, the simulations showed that the optimal foraging area does not maximize population density. Moreover, we found that a polymorphic population emerged, where not a single optimal strategy but a range of optimal strategies can coexist. Using the framework of Adaptive Dynamics, we confirm that the range of optimal areas is not the one that maximizes population size, but the Evolutionary Stable Strategy that can invade a population and not be invaded.
}
\maketitle

\section{Introduction}

 Organisms face a strong selective pressure to optimize their foraging area, driven by a fundamental trade-off between the total foraging area and a limited foraging effort budget \cite{pyke_1984,davis_al_2022,macarthur_pianka_1966}.
 Biotic interactions introduce additional complexity to this adaptive landscape: top predators that prey upon mesopredators might constrain their spatial distribution and impose substantial mortality risk \cite{pringle_al_2019,prugh_al_2009,ritchie_johnson_2009}.
 As individuals adjust their behavior in response to local conditions, ecological tradeoffs might create an apparent tension between organism fitness and population abundance \cite{cabal_al_2020,abrams_2007}.
 Resolving this scale mismatch requires understanding how individual-level mechanisms govern macro-scale ecological outcomes.
 In this context, spatially explicit individual-based models (IBMs) and related analytical approximations are unique tools in allowing the spatial trade-offs to be investigated mechanistically \cite{araujo_home_2010,menezes_al_2025,surendran_al_2025}.

 Classical optimal foraging theory predicts foragers should adopt strategies that maximize net energy intake \cite{pyke_1984}.
 If prey have similar energetic value and require similar handling times, predators should focus on maximizing encounter rates with prey  \cite{charnov_1976}. For animals that typically forage around a central point, as those with a fixed nest or burrow, foraging tends to be more intense around the central point \cite{schoener_1979}.
 While a larger foraging area might increase the total number of potential prey, it decreases the per-prey capture efficiency, modulating the spatiotemporal distribution of interactions.
 Thus, the optimal foraging area emerges from the interplay between local prey spatial distribution and the predator's spatial search range.

 Individual optimization under ecological trade-offs can decouple individual fitness from population abundance, making population size an unreliable proxy for evolutionary optimality \cite{rankin_al_2007}.
 As a result, the optimal foraging strategy in this context might be hard to pinpoint using traditional analytical tools that rely on identifying population-level effects of discrete strategies.
 Nevertheless, strategies might converge to Continuously Stable Strategies (CSS), being at the same time uninvadable (corresponding to an Evolutionarily Stable Strategy, ESS) and evolutionarily accessible via gradual trait changes  \cite{eshel_1983}.
 Furthermore, such a CSS may correspond to a polymorphic equilibrium, wherein different organisms within the population adopt distinct foraging strategies \cite{geritz_al_1998}.
 These theoretical challenges reinforce the importance of spatial, trait-explicit individual-based models, which model individual-level selection directly and allow population-level patterns to emerge mechanistically.

 In a previous study, Araujo \textit{et al.}\cite{araujo_home_2010} proposed an IBM to understand the emergence of optimum foraging area in a predator-prey system.  They consider a bidimensional cellular automata in which the predator can prey in a circular area centered at a focal position. This area can vary over generations, but a trade-off applies: the larger the area, the lower the probability that the predator will prey on a given prey. With a linear trade-off, they showed that the predators evolve towards an optimum area that is not the maximum. The optimum area is the one that balances the effectiveness of predation, which is higher in smaller areas, against access to prey, which increases with greater area.
By assuming that the optimum foraging area maximizes predator population size, the authors could identify a candidate for the optimum area through simulations. They also analytically explored the model by writing the Master Equations within mean-field pair approximations. Although the pair approximations showed good agreement with the simulation for population density, they could not, by the same argument of population size, identify the optimum area.

 In this study, we extend the work of Araujo \textit{et al.}\cite{araujo_home_2010} to a three-trophic food chain, adding a top predator that preys on the first. We also reanalyze the optimal foraging area in terms of individual fitness, rather than maximum population size. This approach revealed that the optimum foraging area is not necessarily the one that maximizes population, but the one that increases the individual fitness.

\section{Model and Analytical approach}

\subsection{The model}

\begin{figure}[!ht]
    \centering
    \includegraphics[width=0.45\linewidth]{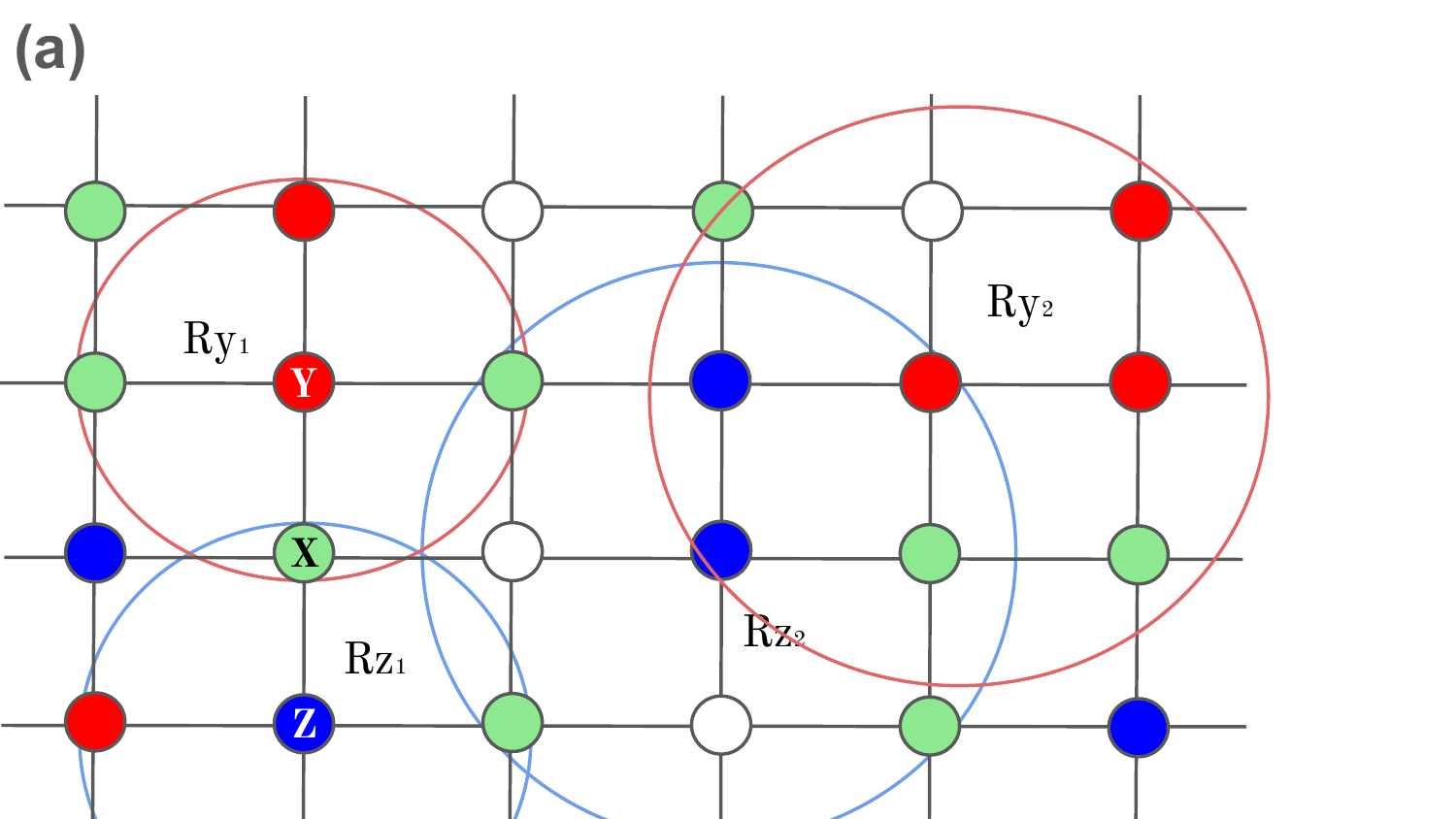}
    \includegraphics[width=0.45\linewidth]{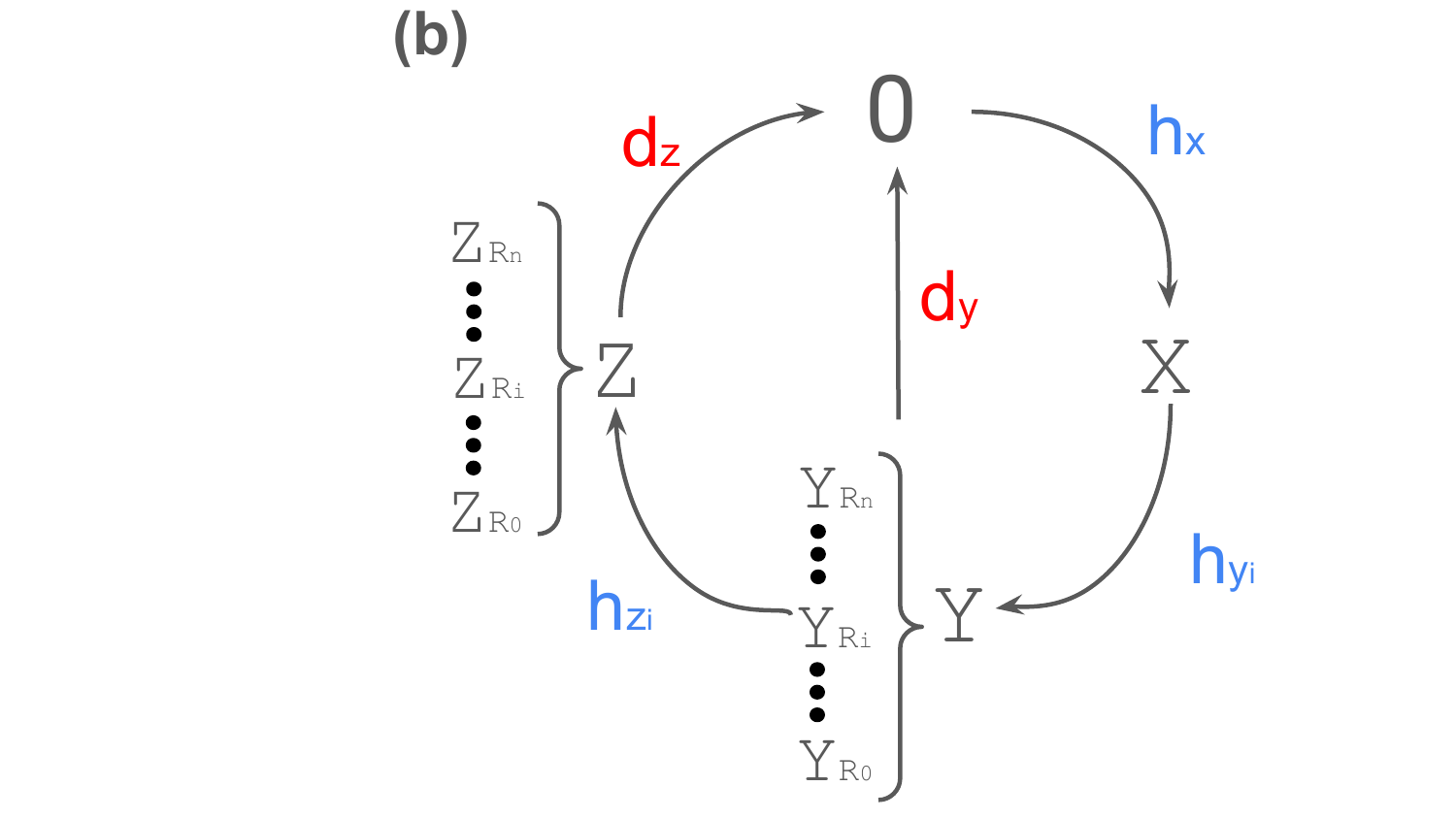}
    \caption{Simplified schematic of the model. (a) Spatial grid with four possible states: empty ($0$), basal prey $X$, mesopredator $Y$, and top predator $Z$. Predators can prey within a circular area centered at their position. (b) State update possibilities: Empty sites can turn into $X$ if there is an $X$ individual in the nearest neighborhood of the site; $X$ individual can be predated by $Y$ individuals if the $X$ site is within the area of foraging defined by the predation radius $R$. Each predator $Y$ may have different radii, so that the subscript $R_0 \dots R_i \dots R_n$ denotes possible radii. $Y$ predators have two possible paths: first, predators can die with probability $d_y$, giving rise to an empty site, or can be predated by $Z$ if there exists a Z predator with a predation radius that encompasses $Y$. The predation radius of $Z$ works in the same way as the predation radius of $Y$. Lastly, $Z$ predators can die with probability $d_z$, returning to an empty site.}
    \label{fig:1}
\end{figure}

The model consists of a three-trophic food chain, spatially organized as a cellular automaton on $L \times L$ sites. Each site can assume one of the four general states: $0$, representing an empty site; $X$, a basal prey; $Y_v$, a mesopredator, which preys on $X$; or $Z_v$, a top predator, which preys on $Y_v$. The subscript $v$ represents the predators' foraging area. The dynamics of each individual depends only on its local neighborhood, in contrast with mean-field models. To reproduce the limitations of time and energy of the predators in search for food, both $Y$ and $Z$ can only prey on sites within a circular area centered at their position. The foraging area is parameterized by the number of sites $v$ a predator can search for food, establishing a trade-off where predation efficiency is inversely proportional to $v$:
\begin{equation}
  h_{A}={g_A}/{v_A}, 
  \label{eq.Global}
\end{equation}
where $g_A$, $A\in\{Y,Z\}$, is a constant parameter that defines the global efficiency of the predator. When a predator successfully preys, the prey dies, and a new predator is born at the predation site. To allow adaptation of the foraging area, the model allows variation of $v$ over individuals and generations: With a mutation probability $\mu$, the foraging area of an offspring can mutate to the immediately larger or smaller neighborhood size relative to its parent (Table \ref{tab:foraging_areas}, Figure \ref{fig:1}).  

\begin{table}[!ht]
\centering
\caption{Predation radii ($R$) and corresponding number of sites ($v$) within the discrete predation neighborhood on a square lattice.}
\label{tab:foraging_areas}
\begin{tabular}{cc@{\hspace{2.5em}}cc@{\hspace{2.5em}}cc}
\toprule
$R$ & $v$ & $R$ & $v$ & $R$ & $v$ \\
\midrule
1.00 & 4  & 3.61 & 44 & 5.39 & 96   \\
1.42 & 8  & 4.00 & 48 & 5.66 & 100  \\
2.00 & 12 & 4.13 & 56 & 5.84 & 108  \\
2.24 & 20 & 4.25 & 60 & \vdots & \vdots  \\
2.83 & 24 & 4.48 & 68 & 10.0 & 316  \\
3.00 & 28 & 5.00 & 80 & \vdots & \vdots  \\
3.17 & 36 & 5.10 & 88 & 50.0 & 7844 \\
\bottomrule
\end{tabular}
\end{table}

\subsubsection*{Model detailed description:}

Each iteration $t$ of the model consists of randomly selecting a site $i$, and updating its state, $\sigma_i (t)$:

\begin{itemize}
    \item  If site $i$ is empty, $\sigma_i (t)=0$ it can turn to a basal prey state $\sigma_i (t+1)=X$.  Each prey $X$  within the four nearest neighborhoods has a probability $h_x$ of reproducing at the site.
    
    \item  If site $i$ is occupied by a prey, $\sigma_i (t)=X$, it can turn a mesopredator  $\sigma_i (t+1)=Y_v$. Each predator $Y$ that has the site $i$ in its predation neighborhood has probability $h_{y_v}$  of changing the state of the site, which represents the birth of one predator $Y$ individual. Prey attacks are prioritized by spatial proximity, with nearer predators acting first (but ensuring that the prey is in its predation area). Attacking order for equally distant predators is randomized. The first successful predation will leave a descendant at site $i$. With probability $\mu$, the predation area of the descendant can mutate: it increases, encompassing the next nearest neighbors (with probability  $\mu/2$) , or decreases, losing the farther nearest neighbors ( with probability $\mu/2$). Consequently, it results in a descendant with lower or higher efficiency in preying at a given site, respectively.  
    
    \item  If site $i$ is occupied by a mesopredator, $\sigma_i (t)=Y_v(t)$ it can transition to an empty site with probability $d_y$, $\sigma_i (t+1)=0$ or to a top predator  $\sigma_i (t+1)=Z_v$. First, the mesopredator dies with probability $d_y$, regardless of the predation area, and the site updates to $\sigma_i (t+1)=0$.  If $Y$ does not die at this stage, similarly to the previous case, each predator $Z$ that has the site $i$ within its predation neighborhood has a probability $h_{z_v}$ of changing the state of the site, which represents the birth of one predator $Z$ individual. As before, the attacks are prioritized by spatial proximity and random among equal distances. The first successful predation will leave a descendant at site $i$. With probability $\mu$, the predation area of the descendant can increase or decrease, resulting in a descendant with lower or higher efficiency in preying at a given site, respectively.

    \item If site $i$ is occupied by an top predator, $\sigma_i (t)=Z_v(t)$ it can only be updated to an empty state, $\sigma_i (t+1)=0$. The probability of $Z_v$ dying is given by $d_z$ regardless of its predation area. 
\end{itemize}
 It is important to note that the foraging area of $Y$ does not affect its chance of being preyed by $Z$, implying that the assumption of greater vulnerability due to a bigger foraging area will not be taken into consideration in this work. However, predators are assumed to allocate more time near their activity center (in the center of the foraging area), granting them priority over closer prey \cite{worton1989kernel,efford2014compensatory}.

\subsection{Analytical approach}
To proceed with an analytical treatment, we follow Araujo \textit{et al.} \cite{araujo_home_2010} and derive the Master Equation associated with the system. To simplify, we assume only two possible foraging areas for each predator, with unbiased mutation between them, and with fixed probabilities $\mu_y$ and $\mu_z$ for mesopredators and top predators, respectively. This results in six possible states for each site: $0$, $X$, $Y_1$, $Y_2$, $Z_1$, and $Z_2$, where the subscript $1$ and $2$ represents the different foraging area for the predators.  The transition probabilities among these states can be written as:

\begin{equation}
T(A\rightarrow B) = 
\begin{cases}
1-(1-h_x)^{N_x} & A = 0, B = X\\
[1-(1-h_{y_1})^{N_{y_1}}(1-h_{y_2})^{N_{y_2}}]f(\mu, h_{y_{1,2}}) & A = X, B = Y_{1,2}\\
d_y & A = Y, B = 0\\
(1-d_y)[1-(1-h_{z_1})^{N_{z_1}}(1-h_{z_2})^{N_{z_2}}]f(\mu, h_{z_{1,2}}) & A = Y, B = Z_{1,2}\\
d_z & A = Z, B = 0
\end{cases}.
\label{transition matrix}
\end{equation}
The term $N_\alpha$ represents the exact count of sites in state $\alpha$ that can interact with the target site within their respective foraging ranges. For instance, the maximum value of $N_x$ is $v_x = 4$, which corresponds to the total number of nearest neighbors an empty site that are basal prey $X$. Similarly, the maximum possible value for $N_{y_1}$ is $v_{y_1}$, corresponding to the full foraging area of mesopredator $Y_1$. While the term $1-(1-h_{y_1})^{N_{y_1}}(1-h_{y_2})^{N_{y_2}}$ represents the probability that at least one mesopredator ($Y_1$ or $Y_2$) successfully predates an $X$ at the focal site, the function $f(\mu, h_i)$ accounts for the probability that the updated state being a given predator $Y_i$:
\begin{equation}
f(\mu, h_i) = \frac{\left[1 - (1 - h_{y_j})^{N_{y_j}}\right]\mu + (1 - \mu)\left[1 - (1 - h_{y_i})^{N_{y_i}}\right]}{2 - (1 - h_{y_i})^{N_{y_i}} - (1 - h_{y_j})^{N_{y_j}}},
    \label{mutation function}
\end{equation}
where index $j$ indicates the other possible predator ($i \neq j$). Therefore, the first term in the numerator of Eq.~(\ref{mutation function}) is the probability of predator $Y_j$ succeeding in predation and a mutation occurring, while the second term is the probability of predator $Y_i$ succeeding in predation and a mutation not occurring. The denominator is the sum of these two probabilities.  

The next step is to write the equations for the mean global densities (occupancy probabilities) of each state over time.  Considering $u$ the percentage of unoccupied (empty, $0$) sites $y_1$, $y_2$, $z_1$ and $z_2$, the percentage of each predator type (Appendix A), we write the general evolution equations for the densities: 

\begin{equation}
    \begin{split}
    x^{t+1} =\; & x^t + u^t \phi_x - x^t \Phi_y, \\\\
    y_i^{t+1} =\; & y_i^t + x^t \psi_y \left[\mu_y \phi_{y_j} + (1 - \mu_y) \phi_{y_i}\right] - y_i^t (1 - d_y) \Phi_z - y_i^t d_y, \\\\
    z_i^{t+1} =\; & z_i^t + (y_1^t + y_2^t)(1 - d_y) \psi_z \left[\mu_z \phi_{z_j} + (1 - \mu_z) \phi_{z_i}\right] - z_i^t d_z, \\\\
    u^{t+1} =\; & 1 - x^{t+1} - y_1^{t+1} - y_2^{t+1} - z_1^{t+1} - z_2^{t+1}.
    \end{split}
    \label{mean_equations}
\end{equation}
where $i$ denotes $1$ or $2$ and $j$ is the complementary of $i$ ($i = 1, j= 2$; $i=2,j = 1$). The functions $\phi$, $\Phi$ and $\psi$ depend on the spatial structure of the system. To evaluate them, we impose spatial closure approximations. Using pair approximations, these functions can be written as:
\begin{equation}
    \begin{split}
    \phi_m(\alpha|\beta) =\; & 1 - \left(1 - h_m q_{\alpha|\beta}\right)^{v_m}, 
    \\\\
    \Phi_p(\alpha_1, \alpha_2|\beta) =\; & 1 - \left(1 - h_{p_2}\right)^{v_{p_2}-v_{p_1}} \left(1 - h_{p_2}q_{\alpha_2|\beta} - h_{p_1}q_{\alpha_1|\beta}\right)^{v_{p_1}}, 
    \\\\
    \psi_p(\alpha_1, \alpha_2|\beta) =\; & \frac{\Phi_p(\alpha_1, \alpha_2|\beta)}{\phi_{p_1}(\alpha_1|\beta) + \phi_{p_2}(\alpha_2|\beta)}, 
    \end{split}
    \label{pa_auxiliary_functions}
\end{equation}
where $q_{\alpha\vert{}\beta} = \frac{\beta\alpha^t}{\beta^t}$ is the conditional probability of observing a state $\alpha$ in the neighborhood of a state $\beta$, derived under the pair approximation. Under the mean-field limit, where spatial correlations are neglected, this conditional probability reduces to the global density, i.e., $q_{\alpha\vert{}\beta} \to \alpha^t$.
For the pair approximation, in addition to the 6 equations corresponding to the single-site population dynamics [Eqs.~(\ref{mean_equations})], we must write the time evolution for each independent pair of states. The pairs involving empty sites ($u$) and occupied states yield:
\begin{equation}
    \begin{split}
    xu^{t+1} =\; & xu^t + d_y(y_1x^t + y_2x^t) + d_z(z_1x^t + z_2x^t) + \\
    & \quad uu^t \phi_x(x|u) - (1-h_x) xu^t \phi_x(x|u) - xu^t \Phi_y(y_1, y_2|x) - xu^t h_x, \\\\
    y_iu^{t+1} =\; & y_iu^t + d_y(y_1y_i^t + y_2y_i^t) + d_z(z_1y_i^t + z_2y_i^t) - (1-d_y) y_iu^t \Phi_z(z_1, z_2|y_i) - \\
    & \quad y_iu^t \phi_x(x|u) + xu^t \psi_y(y_1, y_2|x)\left[\mu_y \phi_{y_{j}}(y_{i}|x) + (1-\mu_y) \phi_{y_i}(y_i|x)\right] - y_iu^t d_y, \\\\
    z_iu^{t+1} =\; & z_iu^t + d_y(z_iy_1^t + z_iy_2^t) + d_z(z_1z_i^t + z_2z_i^t) + \\
    & \quad (1-d_y)\left\{y_1u^t \psi_z(z_1, z_2|y_1)\left[\mu_z \phi_{z_{j}}(z_{j}|y_1) + (1-\mu_z) \phi_{z_i}(z_i|y_1)\right] + \right. \\
    & \quad \left. y_2u^t \psi_z(z_1, z_2|y_2)\left[\mu_z \phi_{z_{j}}(z_{j}|y_2) + (1-\mu_z) \phi_{z_i}(z_i|y_2)\right]\right\} - z_iu^t d_z - z_iu^t \phi_x(x|u).
    \end{split}
    \label{pair_equations}
\end{equation}

The dynamics for a pair of prey and predators ($x y_n$ and $x z_n$) are given by:
\begin{equation}
    \begin{split}
    y_ix^{t+1} =\; & y_ix^t + y_iu^t \phi_x(x|u) - y_ix^t(d_y + h_{y_i}) - (1-d_y) y_ix^t \Phi_z(z_1, z_2|y_i) + \\
    & \quad \left[xx^t - (1 - h_{y_i})y_ix^t\right]\psi_y(y_1, y_2|x)\left[\mu_y \phi_{y_{j}}(y_{j}|x) + (1-\mu_y) \phi_{y_i}(y_i|x)\right], \\\\
    z_ix^{t+1} =\; & z_ix^t + z_iu^t \phi_x(x|u) - z_ix^t d_z - z_ix^t \Phi_y(y_1, y_2|x) + \\
    & \quad (1-d_y)\left\{y_1x^t \psi_z(z_1, z_2|y_1)\left[\mu_z \phi_{z_{j}}(z_{j}|y_1) + (1-\mu_z) \phi_{z_i}(z_i|y_1)\right] + \right. \\
    & \quad \left. y_2x^t \psi_z(z_1, z_2|y_2)\left[\mu_z \phi_{z_{j}}(z_{j}|y_2) + (1-\mu_z) \phi_{z_i}(z_i|y_2)\right]\right\}.
    \end{split}
    \label{pair_prey_equations}
\end{equation}

Finally, the evolution equations for predator-predator pairs are expressed as:
\begin{equation}
    \begin{split}
    y_iz_m^{t+1} =\; & y_iz_m^t + z_mx^t \Phi_y(y_1, y_2|x) + (1-d_y)\Big\{y_iy_1^t \psi_z(z_1, z_2|y_1)\big[\mu_z \phi_{z_{n}}(z_{n}|y_1)  +  \\
    & \quad + (1-\mu_z) \phi_{z_m}(z_m|y_1)\big]+ y_2y_i^t \psi_z(z_1, z_2|y_2)\left[\mu_z \phi_{z_{n}}(z_{n}|y_2) + (1-\mu_z) \phi_{z_m}(z_m|y_2)\right]\Big\} - \\
    & \quad y_iz_m^t(d_y + d_z + h_{z_m}) - y_iz_m^t(1-h_{z_m}) \Phi_z(z_1, z_2|y_n), \\\\
    y_1y_2^{t+1} =\; & y_1y_2^t + y_2x^t(1-h_{y_2}) \Phi_y(y_1, y_2|x) + y_1x^t(1-h_{y_1}) \Phi_y(y_1, y_2|x) + \\
    & \quad \mu_y(xy_2^t h_{y_2} + xy_1^t h_{y_1}) - 2d_y y_1y_2^t - y_1y_2^t \left[\Phi_z(z_1, z_2|y_1) + \Phi_z(z_1, z_2|y_2)\right], \\\\
    z_1z_2^{t+1} =\; & z_1z_2^t + (1-d_y)\left\{(1-h_{z_2})\left[z_2y_1^t \psi_z(z_1, z_2|y_1)\left(\mu_z \phi_{z_2}(z_2|y_1) + (1-\mu_z) \phi_{z_1}(z_1|y_1)\right) + \right.\right. \\
    & \quad \left. z_2y_2^t \psi_z(z_1, z_2|y_2)\left(\mu_z \phi_{z_2}(z_2|y_2) + (1-\mu_z) \phi_{z_1}(z_1|y_2)\right)\right] + \\
    & \quad (1-h_{z_1})\left[z_1y_1^t \psi_z(z_1, z_2|y_1)\left(\mu_z \phi_{z_1}(z_1|y_1) + (1-\mu_z) \phi_{z_2}(z_2|y_1)\right) + \right. \\
    & \quad \left. z_1y_2^t \psi_z(z_1, z_2|y_2)\left(\mu_z \phi_{z_1}(z_1|y_2) + (1-\mu_z) \phi_{z_2}(z_2|y_2)\right)\right] + \\
    & \quad \left. (z_2y_1^t + z_2y_2^t) h_{z_2} \mu_z + (z_1y_1^t + z_1y_2^t) h_{z_1} \mu_z \right\}-2d_zz_1z_2^t,
    \end{split}
    \label{pair_predator_equations}
\end{equation}

The functions $\phi_m(\alpha|\beta)$, $\Phi_p(\alpha_1,\alpha_2|\beta)$ and $\psi_p(\alpha_1,\alpha_2|\beta)$ in the pair evolution equations differ from the definition Eq.~(\ref{pa_auxiliary_functions}) by adding a negative unity on the exponent of these functions, due to the knowledge of one neighbor site.

In addition to the independent pair evolution equations [Eqs.~(\ref{pair_equations})--(\ref{pair_predator_equations})], the remaining auto-pair probabilities are determined by the marginal consistency relations ($\sum_{\alpha} \alpha\beta^t = \beta^t$), enforcing local conservation for each state:

\begin{equation}
    \begin{split}
    uu^t &=\; u^t - ux^t - uy_1^t - uy_2^t - uz_1^t - uz_2^t, \\
    xx^t &=\; x^t - ux^t - xy_1^t - xy_2^t - xz_1^t - xz_2^t, \\
    y_1y_1^t &=\; y_1^t - uy_1^t - xy_1^t - y_1y_2^t - y_1z_1^t - y_1z_2^t, \\
    y_2y_2^t &=\; y_2^t - uy_2^t - xy_2^t - y_1y_2^t - y_2z_1^t - y_2z_2^t, \\
    z_1z_1^t &=\; z_1^t - uz_1^t - xz_1^t - y_1z_1^t - y_2z_1^t - z_1z_2^t, \\
    z_2z_2^t &=\; z_2^t - uz_2^t - xz_2^t - y_1z_2^t - y_2z_2^t - z_1z_2^t,
    \end{split}
    \label{pair_conservation_relations}
\end{equation}
where the spatial symmetry condition $\alpha\beta^t = \beta\alpha^t$ applies to all pairs.

\subsection{Investigated scenarios}

We first investigated the dynamics for the three-trophic food chain without mutation ($\mu_y=\mu_z=0$), and compared the simulation to the mean-field approximations and pair approximation.  Next, we examined scenarios in which only predators at a particular trophic level were capable of mutation (either only $\mu_y>0$ or only $\mu_z>0$).
Finally, we investigated the scenario where both mesopredators and top predators could mutate ($m_y>0 \,\land \,\mu_z>0$). The individual fitness was evaluated using an invasion scenario in which a mutant variation in foraging area is introduced into a resident population \cite{SMITH1973}.  

For all simulations, we fixed the grid size as a square lattice of $100\times100$ cells, and each simulation was run over at least $\times10^4$ generations, where each generation corresponds to a $100\times100=10^4$ iterations, i.e. a Monte Carlo Step (MCS). The results present the convergent mean densities of the populations after a transient of $5\times10^3$ generations. 
For the simulations, the results were also averaged over $10$ independent realizations. The probability of $X$ colonizing a first-neighbor unoccupied site was fixed at $h_x=0.5$, and the mortality of predators at $d_y=0.15$ and $dz=0.2$. When the mutation was allowed, it was fixed at $\mu=0.05$. The general effect of the global efficiency, $g_y$ and $g_z$ (Eq.~(\ref{eq.Global}) on the dynamics was investigated between the range $]0,4]$. %

\section{Results and Discussion}

\begin{figure}[!ht]
 \centering
\includegraphics[width=0.95\linewidth]{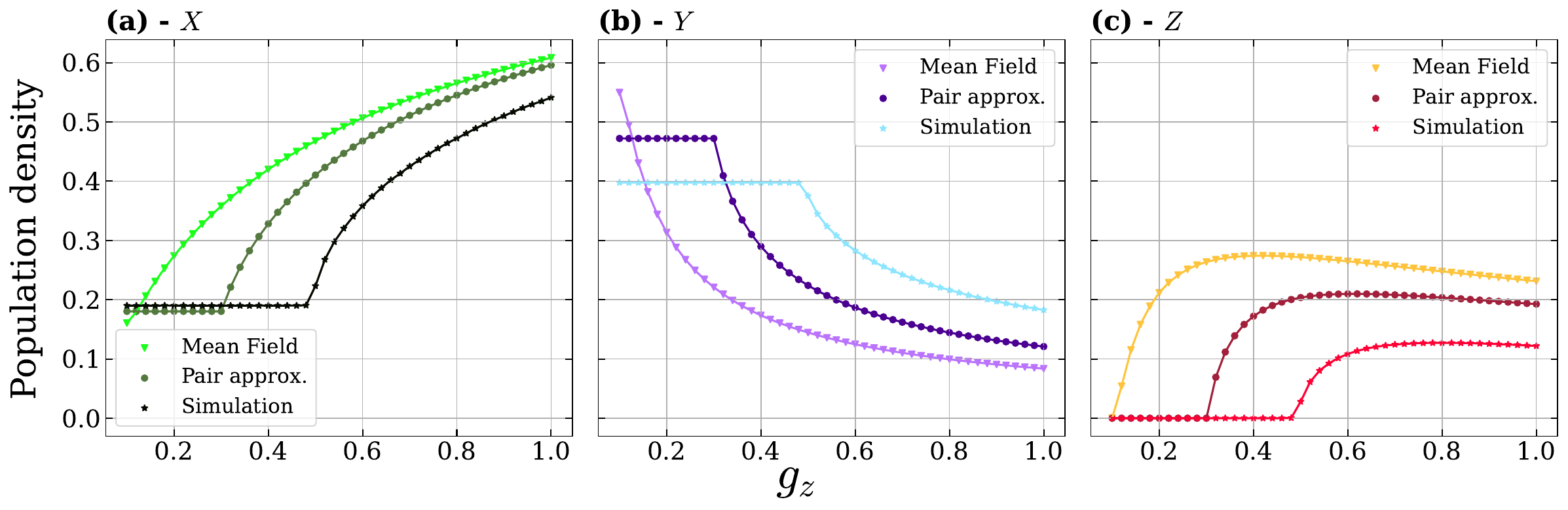}
    \caption{Comparison between simulations, mean-field and pair approximations for the three-trophic food chain without mutation. The graphs show the convergent mean density after a transient of $10^4$ MCS of (a) prey population $X$, (b) mesopredator $Y$, and (c) top predator $Z$ as a function of $Z$ efficiency in predation, $h_z$. For these graphs $v_y=v_z=4$, $g_y=1.2$ (which results in $h_y=0.3$).}
    \label{fig:2}
\end{figure}

Figure (\ref{fig:2}) presents the convergent mean densities of the three-trophic food chain—prey ($X$), mesopredator ($Y$), and top predator ($Z$)—as a function of the top predator’s predation efficiency ($h_z$), comparing simulations, pair, and mean-field approximations. As $h_z$ increases, the system transitions from a two-trophic state (where $Z$ is absent) to a full three-trophic coexistence state. The simulations show that the top predator $Z$ emerges only above a critical threshold $h_z\approx 0.5$, exhibiting a non-monotonic behavior characterized by a rapid initial growth followed by a gradual decline at high efficiency values. This decline in $Z$ relaxes the top-down pressure on the mesopredator $Y$, which in turn causes the basal prey density $X$ to monotonically increase with $h_z$. The superior agreement of the pair approximation over the mean-field approach highlights the presence of spatial inhomogeneities and local clustering in the system.  In particular, the pair approximation identifies the existence of a critical threshold for top predator persistence, which is not present in the mean-field approximation.

\begin{figure}[!ht]
\centering
    \includegraphics[width=1.0\linewidth]{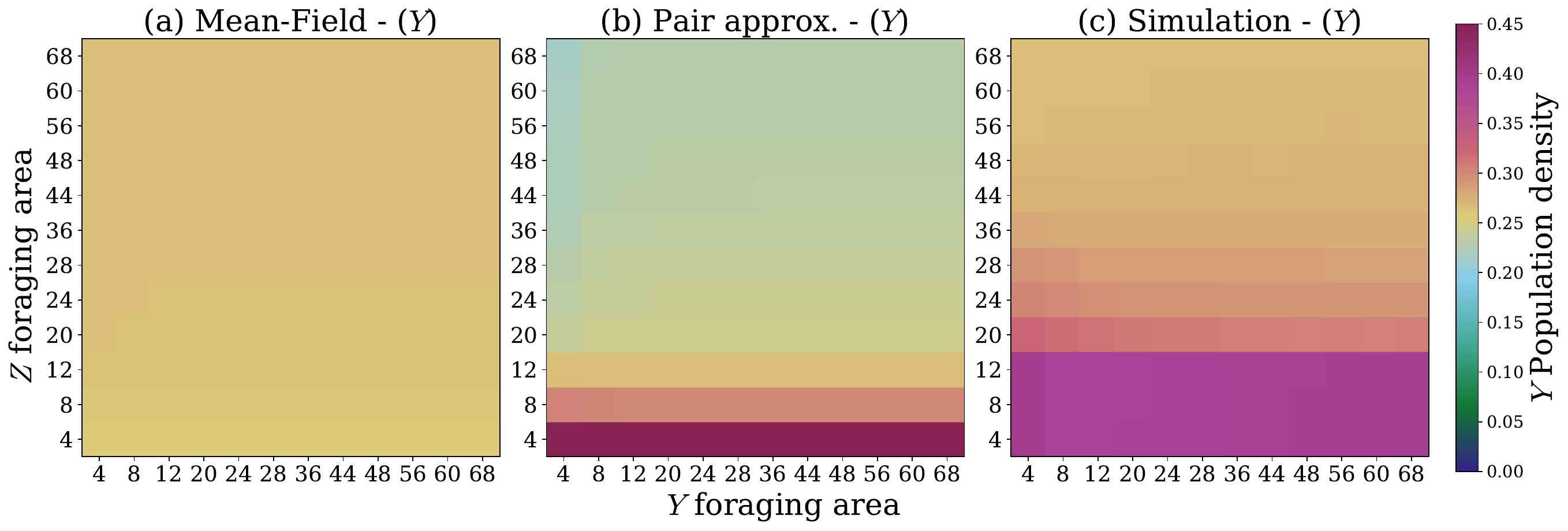}
    \includegraphics[width=1.0\linewidth]{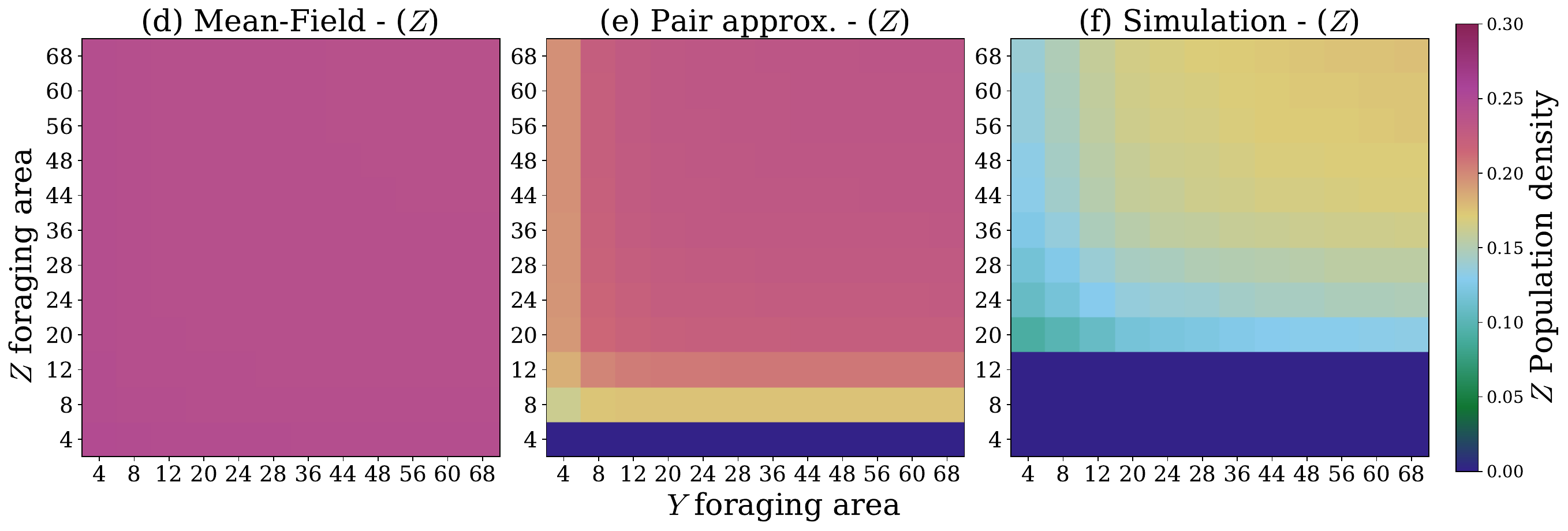}
    \caption {Comparison between simulations, mean-field and pair approximations for the three-trophic food chain without mutation. The first row shows the mean density of mesopredator $Y$, and the second row the density of top predator $Z$, both as a function of the foraging areas.} 
    \label{fig:3}
\end{figure}

The better qualitative agreement of the pair approximation over the mean field can also be evidenced for different foraging areas (Fig. \ref{fig:3}): while the mean-field approximation practically does not predict variation of population size under different foraging areas, the pair approximation correctly captures how predator density changes with foraging area, although quantitative agreement is lacking. Importantly, the pair approximation correctly predicts the existence of a minimum threshold for $Z$ foraging area, below which $Z$ goes extinct locally. The pair approximation, as well as the simulations, show that the density of $Y$ is almost insensitive to its foraging area, but is more sensitive to $Z$'s foraging area: the greater the $Z$ foraging area, the lower the $Y$ population. It is interesting to note that there is no local maximum for $Y$ or $Z$ population densities (Fig. \ref{fig:3}). So, in a naive way, one could conclude that there is not an optimal foraging area for $Y$, and it would be as large as possible for $Z$. 

\begin{figure} [!ht]
    \centering
    \includegraphics[width=1\linewidth]{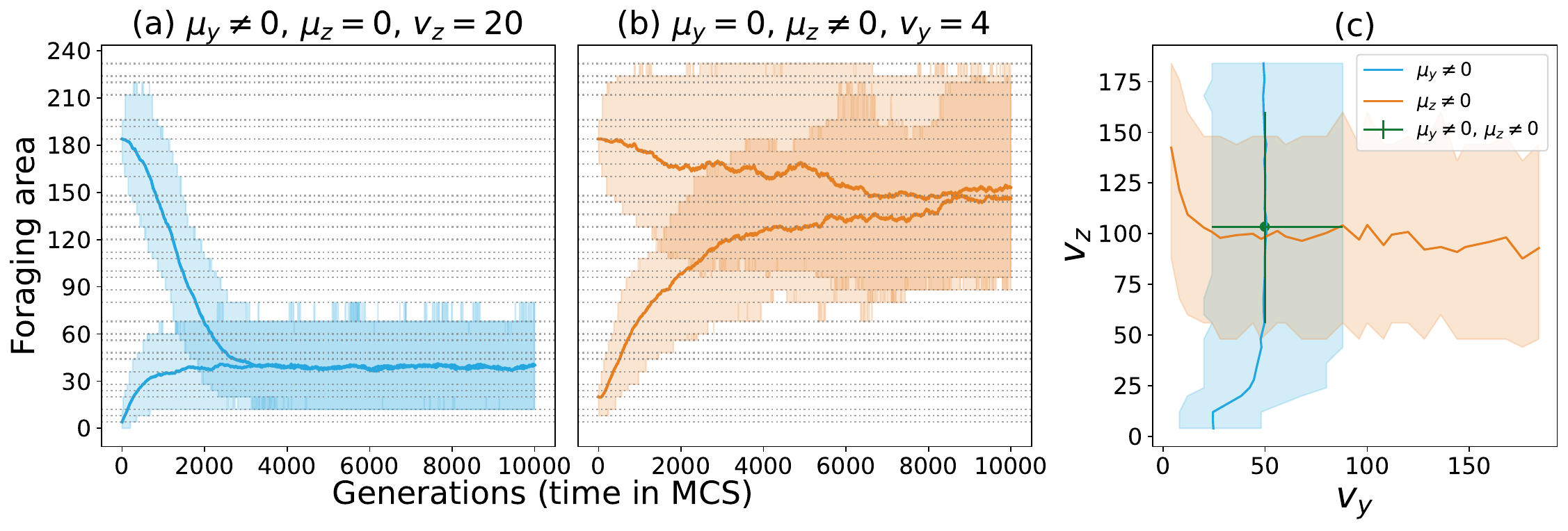}
    \caption{
    Simulation of a three-trophic food chain under foraging area mutation. (a) Time evolution of foraging area for the mesopredator ($Y$) when only $Y$ can mutate ($\mu_y = 0.05$, $\mu_z = 0$, and $v_z = 20$), averaged over five simulations for each of two distinct initial conditions. (b) Time evolution of the mean foraging area for the top predator ($Z$) when only $Z$ can mutate ($\mu_y = 0$, $\mu_z = 0.05$, and $v_y = 4$), also averaged over five replicates for two different initial conditions. (c) Stationary mean foraging area (after $2 \times 10^5$ MCS for 5 replicates) for $Y$ (blue, where $\mu_y = 0.05$, $\mu_z = 0$) and $Z$ (orange, where $\mu_y = 0$, $\mu_z = 0.05$) as a function of the fixed foraging area of the other predator. Shaded regions represent 90\% percentile intervals. The green point with error bars denotes the equilibrium values and 90\% percentile intervals for the scenario where both species can mutate ($\mu_y=0.05, \mu_z=0.05$). 
    }
    \label{fig:4}
\end{figure}
\begin{figure}[!ht]
    \centering
   \includegraphics[width=1.0\linewidth]{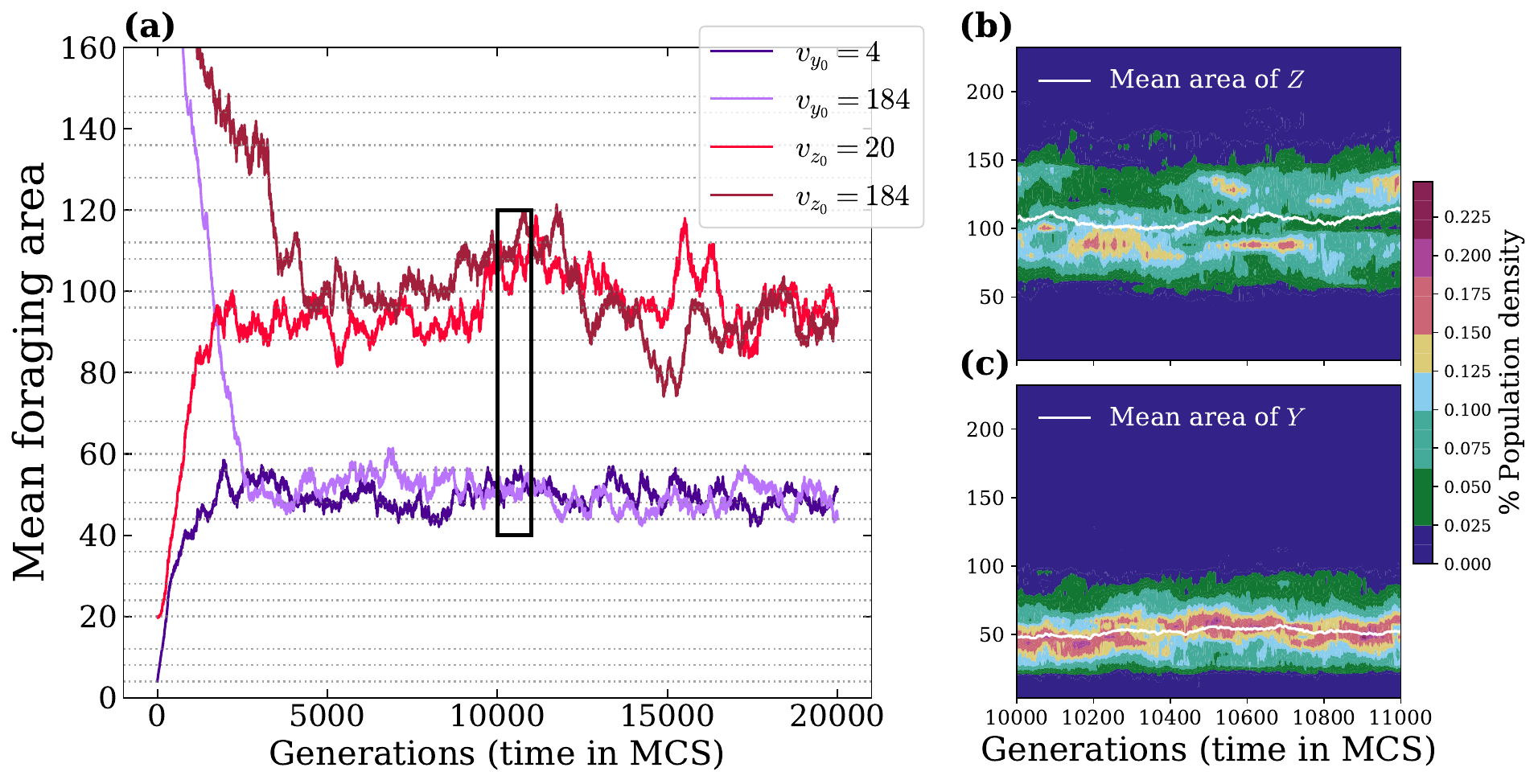}
    \caption{(a) Temporal evolution of foraging area of mesopredator ($ v_y$) and top predator ($ v_z$) when both mutation rates are nonzero for two distinct initial conditions. Compared with cases in which only one species can mutate, $Y$ predators reach a stationary foraging area at higher values (44 to 56 sites), while $Z$ predators establish smaller foraging areas (94 to 108 sites). (b) $Z$ and (c) $Y$ population frequency of foraging areas over time in the interval $[1.0\times10^4, 1.1\times10^4]$.}
    \label{fig:6}
\end{figure}

This naive expectation is refuted when evolutionary dynamics are explicitly incorporated into the model, resulting in convergence toward an optimal foraging area regardless of the initial conditions (Figures \ref{fig:4}a-b and \ref{fig:6}a). To systematically investigate these adaptive mechanisms, we first tested scenarios where mutation was restricted to a single trophic level. When only the mesopredator's foraging area mutates, it evolves to a constant value as long as $Z$'s foraging range satisfies $v_z \gtrsim 40$ (Figure \ref{fig:4}c, blue line). The foraging area for $v_z \leq 12$ converges to lower values due to the extinction of $Z$ predator, as shown in Figure ~(\ref{fig:3}f). Similarly, $Z$'s optimal foraging area plateaus and converges to a constant value once $Y$'s foraging area exceeds a critical size, $v_y \gtrapprox 25$ (Figure \ref{fig:4}c, orange line). From Figure (\ref{fig:3}) it can be noticed that for bigger foraging area, both $Y$ and $Z$ populations do not change significantly, a factor that might influence the convergence seen on Figure \ref{fig:4}c for higher area values of the population that does not mutate. 

The individual's foraging area does not converge to a unique value, revealing that rather than a single optimal foraging area, the spatial dynamics favor a bounded region of viable strategies (Figure \ref{fig:6}b-c). Within this domain, distinct lineages characterized by different foraging areas continuously emerge, coexist, and vanish over generations. This polymorphism suggests that local spatial fluctuations dynamically favor individuals with higher search efficiency (smaller foraging area) at times, and those with a broader search range (larger foraging area) at others. Remarkably, both strategies can even coexist within the population (see, for example, Figure \ref{fig:6}c between generations 10800 and 11000).
 
\begin{figure}[!ht]
    \centering
     \includegraphics[width=1\linewidth]{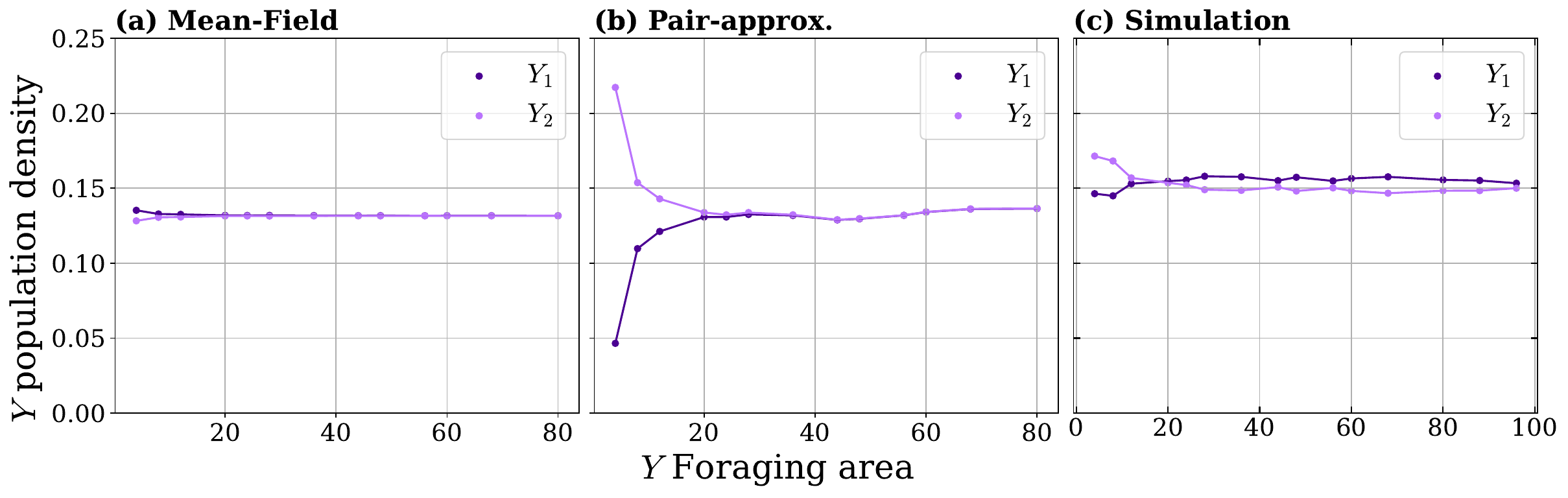}
    \includegraphics[width=1\linewidth]{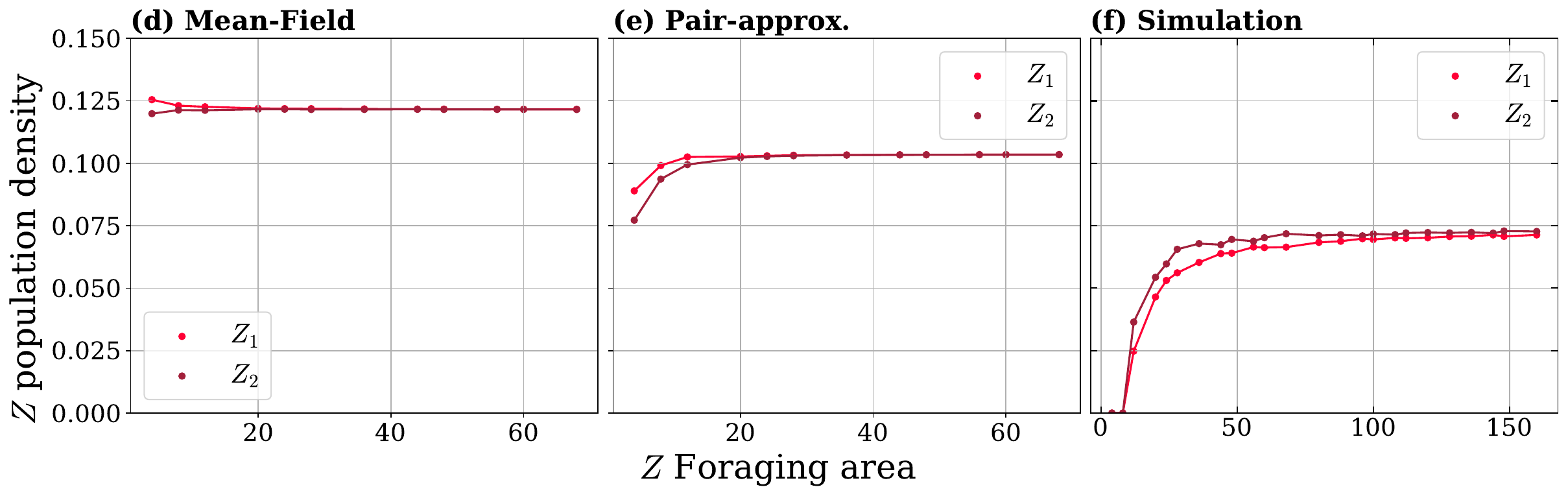}
   \caption{Evolution of predator population densities under two-state mutation competing scenarios. The top row shows the mean population density of the mesopredator $Y$ when only $Y$ is subject to mutation, while the bottom row shows the top predator $Z$ when only $Z$ can mutate, both as a function of the baseline foraging area $v_i$. Predictions from the mean-field approximation (a,d) and pair approximation (b,e) are compared against full stochastic simulations (c,f). Foraging areas under mutation correspond to competing adjacent discrete pairs $(v_i, v_{i+2})$, such as $(4, 12)$, $(8, 20)$, $(24, 36)$, and so forth.}
\label{fig:5}
\end{figure}
 A much simplified version of the system allows mutation only between two competing foraging areas. This simplification makes the analytical approach feasible, via Eqs.~(\ref{mean_equations}) and (\ref{pair_equations}). Restricting the mutation to a single trophic level while keeping the foraging area of the other level fixed at different values, we compare how well the approximations describe the simulations (Figure \ref{fig:5}). When mutation occurs only for the mesopredator $Y$ (Figure \ref{fig:5}, first row), the mean-field prediction yields almost identical population densities for both $Y_1$ and $Y_2$, with a slightly higher value for $Y_2$ at small foraging areas: $(v_1=4, v_2=12)$ and $(v_1=8, v_2=20)$. For these small foraging areas, the pair approximation predicts a much larger difference in population density and, more importantly, an opposite trend: a smaller foraging area results in a smaller population size until they merge around $v_1=20$ (where $v_2=36$). This smaller population density of $Y_1$ is also observed in the stochastic simulations, revealing the presence of spatial organization. However, for larger foraging areas, the simulation does not show a merging of population densities, but rather an inversion where the density of $Y_1$ overcomes $Y_2$.

 When mutation is applied only to the top predator $Z$ (Figure \ref{fig:5}, second row), the extinction of $Z$ for small foraging areas is observed ($v_1 < 24$, corresponding to $v_2 < 36$). Beyond this threshold, $Z$'s density grows rapidly and levels off into a plateau. Although neither analytical approximation can predict the extinction event, the pair approximation reproduces the general trend of population recovery, whereas the mean-field approximation remains insensitive to variations in the foraging area.
 
 The inversion of population density of mesopredators observed in the simulations (Figure \ref{fig:5}c) reveals that a larger foraging area favors population density up to a certain value and then a smaller one maximizes it. This inversion occurs at $v_1=20$, suggesting it as a good candidate for the optimum foraging radius \cite{araujo_home_2010}. However, when we compare to the scenario where the mutation is not restricted to two competing areas, we see that the optimum foraging area is not around $20$, but $44$ (Figure~\ref{fig:4}). This further indicates that optimal strategies derived for simplified models and based on abundance criteria might be insufficient to capture accurate system behavior, particularly for polymorphic asymptotic states.

\begin{figure}[!ht]
    \centering
    \includegraphics[width=1\linewidth]{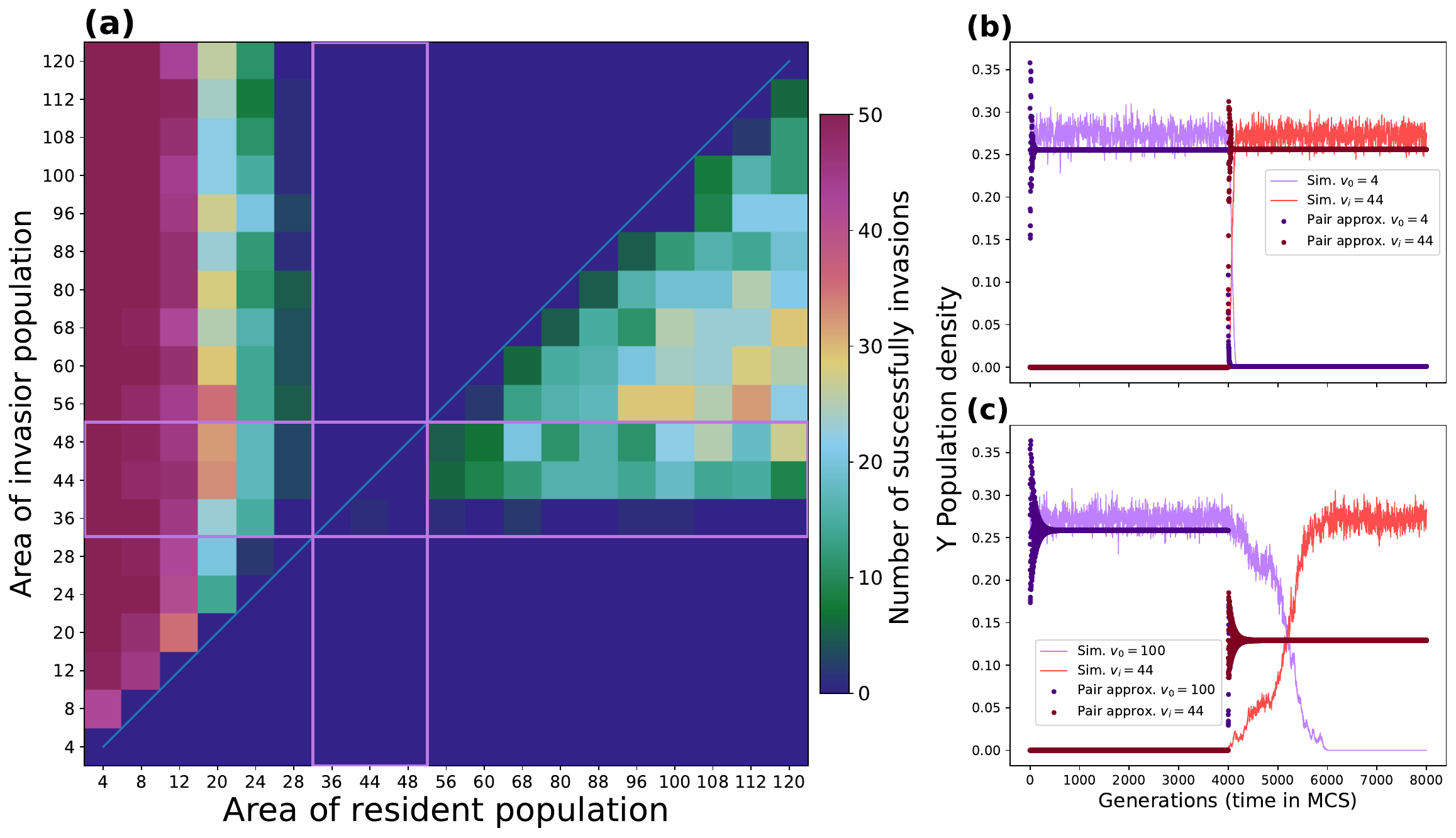}
    \caption{Invasion dynamics of a mutant mesopredator ($Y$) in a three-trophic system without ongoing mutation. The system evolves for $5000$ generations with a fixed resident foraging area $v_{0}$, after which a single mutation event introduces an invader mutant ($Y_{i}$) with foraging area $v_{inv}$ at $1\%$ of the resident population size. (a) Pairwise invasibility heat map obtained from stochastic simulations, showing the percentage of successful invasions (out of 50 replicates) within $5000$ post-mutation generations as a function of the resident ($v_{0}$) and mutant ($v_{i}$) foraging areas. The diagonal line highlights $v_{0}=v_{i}$ while vertical and horizontal lines highlight the range of non-invaded foraging areas and their capacity to invade. (b, c) Time series of representative invasion scenarios showing the population densities of resident (purple) and invader (red) predators obtained from stochastic simulations and pair approximation predictions for different values of foraging areas. 
    For these graphs For these graphs $v_z = 48$ was fixed.}
     \label{fig:7}
\end{figure}

To evaluate whether the optimal foraging strategy constitutes an Evolutionarily Stable Strategy (ESS) in the framework of Adaptive Dynamics \cite{hofbauer1998evolutionary}, we performed invasion experiments (Figure \ref{fig:7}). The three-trophic system was allowed to reach a stationary state without mutation over $5000$ generations. At generation $5000$, a single mutation event occurred, introducing an invader mutant population with a distinct foraging area at a low initial frequency ($1\%$ of the resident population). The simulations showed that when the resident population has a very small foraging area ($\lesssim 24$), any mutant with a larger foraging area successfully invades and the resident population goes extinct(Figure \ref{fig:7}a). Conversely, when the resident population exhibits an excessively large foraging range ($\gtrsim 80$), mutant invaders with smaller, more efficient foraging areas are favored to invade and promote the extinction of the residents. When the resident population has a foraging area between $36$ and $48$, no invasion occurs, leading to its extinction in a few generations. Interestingly, this holds true even when the foraging area of the invader is between this range.   This result may not seem to be in line with the result that reveals that multiple lineages can coexist (Figure \ref{fig:6}b,c). However, it is important to note that in the scenario of Figure (\ref{fig:6}), mutation is constant over time, which may favor lineage coexistence. Another interesting point to highlight is that the successful invader population's density does not overcome that of the resident population (Figure \ref{fig:7}b,c) and, as the resident and invader do not coexist, the non-extinct population density evolves to the values observed in Figure (\ref{fig:3}). Again, looking only at population density (Figure (\ref{fig:3}), it is not possible to infer the range of optimal foraging areas. 

The pair approximation also predicted that a resident population can be invaded when it has a small foraging area ($v_0$=4); a larger foraging area can invade this population and cause it to go extinct (Figure \ref{fig:7}b). However, in contrast to stochastic simulations, under the pair approximation there is no foraging area that prevents the growth of the invader population, and both populations can coexist (Figure \ref{fig:7}c).

\section{Final Remarks}
In this study, we investigated the evolutionary dynamics of a three-trophic food chain where the predation foraging area can evolve (mutate) over generations. We considered a trade-off between the predator foraging area and its efficiency of preying on a given prey (Eq. \ref{eq.Global}). We implemented this model using a stochastic cellular-automata model and also developed mean-field and pair approximations. We observed that the foraging areas of both intermediate and top predators reliably converge towards a limited range of values regardless of initial conditions. Rather than fixating on a single isolated value, the evolutionary dynamics sustain a bounded range of viable strategies that allow continuous polymorphic coexistence and turnover among competing lineages.

We did not observe an alignment between population size and individual fitness for the optimal foraging areas recorded (compare abundances on Figure~\ref{fig:3} and optimal foraging areas from Figure~\ref{fig:4}).
While larger population abundances were reached for arbitrarily large foraging areas (Figure~\ref{fig:3}), the emerging CSS strategy covered a much more limited range ($v_y \in [20, 100]$ and $v_z \in [50, 150]$, Figure~\ref{fig:4}c and \ref{fig:5}b-c), substantially smaller than the theoretical maximum foraging area allowed by the system size ($v_{\text{max}} \approx 10^4$).
In a hypothetical well-mixed scenario achieved through exceptionally large foraging ranges ($v_y=v_z=v_{\text{max}}$) any predator would have a vanishingly small chance of consuming any prey, even when accounting for priority effects due to distance.
As such, any mutation that reduced foraging area would increase predation (and reproduction) probabilities, increasing individual fitness at the expense of reducing the number of available prey for the resident population.
Thus, what we observed in this system is akin to the ecological tragedy of the commons rooted in the spatial structure of preys and predators \cite{rankin_al_2007,cabal_al_2020}.

Subpopulations with very small foraging area also risk being outcompeted as they become extremely susceptible to local depletion of prey \cite{melbourne_hastings_2008}.
Therefore, some intermediate value of foraging area optimizes individual fitness, but it is not clear which foraging area that should be without the precise model-based investigation developed here.
Interestingly, our analyses revealed that a range of foraging ranges that coexisted in simulations, indicating that no single foraging area can dominate the system (Figures~\ref{fig:6} and \ref{fig:7}), i.e. no stable dominant monomorphic strategy.
This is further demonstrated by the invasibility analysis for the mesopredator (Figure~\ref{fig:7}).
For instance, a resident population with foraging area 36 could not be invaded by populations with different foraging areas and thus corresponds to ESS.
However, the same population was not able to invade resident populations with larger foraging areas, failing to be a monomorphic CSS.

The balance of invasibility among multiple strategies and the continuous emergence of new foraging areas through mutation led to a persistent diversity of foraging areas in our simulations (Figures~\ref{fig:4},\ref{fig:6} and \ref{fig:7}).
Considered together, our results suggest that the mixed strategy of multiple foraging areas is a CSS.
The diversity of foraging areas we observed is mirrored in extensive empirical data \cite{medici_al_2022,morato_al_2018}. 
However, the intrinsic complexity of this dynamical mixing of multiple strategies hindered analytical investigations.

Analytically, the pair approximation offers a substantial improvement over standard mean-field theory by incorporating local spatial correlations and reproducing the qualitative dynamics of population densities (Figures \ref{fig:2}, \ref{fig:3}, \ref{fig:5}), a result widely observed in the literature \cite{tome_stable_2007, Gleeson2013, NAKAMARU1997}. However, it fails to predict the optimal foraging area identified in stochastic simulations. In our analytical invasion dynamics analyses (Figure \ref{fig:7}b-c), larger foraging areas can readily invade or coexist with smaller ones, but the pair approximation fails to produce a bounded strategy range that is completely uninvadable by any mutant population. This discrepancy may be due to the priority sequence of predation imposed in the simulations and not explicitly described by the approximations. In the simulation, the closer predators have priority in preying, and successful predation results in an offspring with the same foraging area (plus a probability of mutation) as its parent. Thus, an enlarged foraging area reduces capture efficiency for nearby prey while offering no priority over distant ones (which has supporting evidence \cite{worton1989kernel,efford2014compensatory}). Such distance-dependent effects can generate spatial heterogeneity in search efficiency, thereby influencing target encounter rates, consistent with theoretical studies showing that movement strategies and environmental heterogeneity can strongly affect the efficiency of random search \cite{viswanathan_optimizing_1999,menon_random_2024,martinez-garcia_al_2020}.

\section*{Declarations}
\subsection*{Funding}
RM was funded, in part, by the São Paulo Research Foundation (FAPESP), Brasil ( \#2024/18255-0 and \#2021/14335-0, ICTP-SAIFR). MAMA and SBLA were funded by  Conselho Nacional de Desenvolvimento Científico e Tecnológico (CNPq) for support (Grants \#303814/2023-3 and \#3304094/2025-0, respectively). 

\section*{Acknowledgments}
LMP and SBLA acknowledge computational support from Professor Carlos M. de Carvalho at LFTC-DFis-UFPR.

\appendix
\section*{Appendix A}

In this appendix, the process of evaluating the averages present in the Master Equation is discussed in greater detail. The development of the approximations was first made for two trophic levels in \cite{araujo_home_2010}. In this appendix, we generalized to the presence of top predator $Z$. Starting from the discrete Master equation
\begin{equation}
        P(\sigma,t+1) = P(\sigma, t) + \sum_m\sum_{i=0}^{N}\left[P(^m\rho_i,t)T(^m\rho_i\rightarrow\rho)-P(\rho_i,t)T(\rho_i\rightarrow\rho^m_i) \right],
\end{equation}
Where $\sigma$ denotes the state of the system, given by the individual state of each site, $\sigma = \sigma(\rho_1,\rho_2,\dots,\rho_i,\dots,\rho_N)$ with $\rho = \{u, x, y_i,z_i\}$. The population average of each species at a given time can be obtained with the use of Kronecker delta $\delta(\alpha,\beta)$ in conjunction with Master Equation $\langle\omega(t)\rangle = \sum_i \left[P(\sigma_i, t)\delta(\sigma_i,\omega)\right]$, since each site is occupied by only one individual. In this context, $\omega$ denotes one of the possible states of a site, i.e, $\{u, x, y_i,z_i\}$:
\begin{equation}
    \langle \omega (t+1) \rangle = \langle \omega(t) \rangle + \sum_{m} \langle \delta({}^i\sigma_m, \omega) T(\sigma_i \to {}^i\sigma_w) \rangle - \sum_{m} \langle \delta(\sigma_i, \omega) T(\sigma_i \to {}^i\sigma_m) \rangle,
\end{equation}
where the transition matrix is given by  Eq.~(\ref{transition matrix}). Here we will show how to calculate the three more complicated averages that come from Eq.~(\ref{transition matrix}):
\subsection{$\langle\delta(\sigma_i,\alpha)\delta(\sigma_j,\theta)[1-(1-h_\beta)^{v_\beta}]\rangle$}

For this term, we proceed expanding $(1-h_\beta)^{N_\beta}$ in a Taylor series around $h_\beta = 0$:
\begin{equation}
\begin{split}
&\langle\delta(\sigma_i,\alpha)\delta(\sigma_j,\theta)[1-(1-h_\beta)^{N_\beta}]\rangle = h_\beta \langle\delta(\sigma_j,\theta) \delta(\sigma_i, \alpha) N_\beta \rangle 
\\
&-\frac{h_\beta^2}{2!} \langle \delta(\sigma_j,\theta)\delta(\sigma_i, \alpha) N_\beta (N_\beta - 1) \rangle 
+\frac{h_\beta^3}{3!} \langle \delta(\sigma_j,\theta)\delta(\sigma_i, \alpha) N_\beta (N_\beta - 1)(N_\beta - 2) \rangle \dots
\end{split}
\end{equation}
Using the fact that the number of neighbors of $\alpha$ that can interact with it is given by $N_\beta$ in the site $l$, one can write 
\begin{equation}
\langle N_\beta\delta(\sigma_i,\alpha)\delta(\sigma_j,\theta)\rangle=\sum_{l=1}^{v_\beta}\langle\delta(\sigma_i,\alpha)\delta(\sigma_j,\theta)\delta(\sigma_l,\beta)\rangle = \sum_{l=1}^{v_{\beta}}P_{i,j,l}(\alpha\theta\beta).
\label{n_vizinhos}
\end{equation}
Quadratic or high-order terms in $N_\beta$ may be separated in the cases in which the site of both sums is the same. In this case, one can proceed as follows:
\begin{equation}
\begin{split}
    &\langle N_\beta(N_\beta-1)\delta(\sigma_i,\alpha)\delta(\sigma_j,\theta)\rangle = \sum_{l\neq k}\langle\delta(\sigma_l,\beta)\delta(\sigma_k,\beta) \delta(\sigma_i,\alpha)\delta(\sigma_j,\theta) \rangle+\sum_{l=k}\langle\delta(\sigma_l,\beta)\delta(\sigma_i,\alpha)\delta(\sigma_j,\theta)\rangle -\\
    &\sum_{l}\langle\delta(\sigma_l,\beta)\delta(\sigma_i, \alpha)\delta(\sigma_j,\theta)\rangle\\
    & = \sum_{l\neq k} \langle\delta(\sigma_l,\beta) \delta(\sigma_k,\beta) \delta(\sigma_i,\alpha) \delta(\sigma_j,\theta) \rangle = \sum_{l\neq k}^{v_\beta}P_{i,j,k,l}(\alpha\theta\beta\beta)
\end{split}
\end{equation}

Substituting the other terms and simplifying the common terms, we obtain
\begin{equation}
\begin{split}
&\langle \delta(\sigma_l, \theta)\delta(\sigma_i, \alpha)[1 - (1 - h_\beta)^{N_\beta}] \rangle = h_\beta \sum_{l}^{v_\beta} P_{l,i,j}(\alpha\theta\beta)\\
&- \frac{h_\beta^2}{2!} \sum_{l \neq k}^{v_\beta} P_{l,i,j,k}(\alpha\theta\beta\beta) + \frac{h_\beta^3}{3!} \sum_{l \neq k \neq m}^{v_\beta} P_{l,i,j,k,m}(\alpha\theta\beta\beta\beta) - \dots
\end{split}
\label{media_intermediaria}
\end{equation}
Until this far, all expressions are exact. To proceed, one could use one of the approximations given below:

\begin{equation}
\begin{split}
    &P(\alpha\beta\gamma, t) \stackrel{pair}{\approx} \frac{P(\alpha\beta,t)P(\alpha\gamma,t)}{P(\alpha)}\\
    &P(\alpha\beta, t) \stackrel{mf}{\approx} P(\alpha, t)P(\beta, t).
\end{split}
\label{approxim.}
\end{equation}
Under pair-approximation, for example, one can identify the expression Eq.~(\ref{media_intermediaria}) as the series of
\begin{equation}
\begin{split}
&\langle \delta(\sigma_l, \theta)\delta(\sigma_i, \alpha)[1 - (1 - h_\beta)^{v_\beta}] \rangle
    \approx \alpha\theta^t \left[ 1 - \left( 1 - \frac{h_\beta (\alpha\beta^t)}{\alpha^t} \right)^{v_\beta - 1} \right],
\end{split}
\label{media_1_resultado}
\end{equation}
corresponding to the first equation in Eq.~(\ref{pa_auxiliary_functions}).

\subsection{$\langle\delta(\sigma_i,\alpha)\delta(\sigma_j,\theta)[1-(1-h_{\beta_1})^{N_{\beta_1}}(1-h_{\beta_2})^{N_{\beta_2}}]\rangle$}
Similarly to the previous subsection, to calculate this average we expand both $h_{\beta_1}$ and $h_{\beta_2}$ around $h_\beta = 0$. By combining terms of equal order in $h_\beta$, and using Eq.~(\ref{n_vizinhos})
\begin{equation}
    \begin{split}
        &\langle\delta(\sigma_i,\alpha)\delta(\sigma_j,\theta)[1-(1-h_{\beta_1})^{N_{\beta_1}}(1-h_{\beta_2})^{N_{\beta_2}}]\rangle = \langle\delta(\sigma_i,\alpha)\delta(\sigma_j,\theta)I\rangle+\langle\delta(\sigma_i,\alpha)\delta(\sigma_j,\theta)II\rangle\\
        &+\langle\delta(\sigma_i,\alpha)\delta(\sigma_j,\theta)III\rangle,\\
        &I = h_{\beta_1}N_{\beta_1}+h_{\beta_2}N_{\beta_2}\\
        &II = h_{\beta_1}h_{\beta_2}N_{\beta_1}v_{\beta_2}-\frac{h^2_{\beta_1}}{2}N_{\beta_1}(N_{\beta_1}-1)-\frac{h^2_{\beta_2}}{2}N_{\beta_2}(N_{\beta_2}-1)\\
        &III = N_{\beta_1} N_{\beta_2} (N_{\beta_2} - 1) \frac{h_{\beta_1} h_{\beta_2}^2}{2} + N_{\beta_1} (N_{\beta_1} - 1) N_{\beta_2} \frac{h_{\beta_1}^2 h_{\beta_2}}{2} + N_{\beta_1} (N_{\beta_1} - 1) (N_{\beta_1} - 2) \frac{h_{\beta_1}^3}{6}\\
        &+N_{\beta_2} (N_{\beta_2} - 1) (N_{\beta_2} - 2) \frac{h_{\beta_2}^3}{6}.
    \end{split}
    \label{media_2}
\end{equation}

To deal with the sums terms of Eq.~(\ref{media_2}) one can use the linear condition:
\begin{equation}
\langle\delta(\sigma_i,\alpha)\delta(\sigma_j,\theta)[N_{\beta_1}h_{\beta_1}+N_{\beta_2}h_{\beta_2}]\rangle = \langle\delta(\sigma_i,\alpha)\delta(\sigma_j,\theta)N_{\beta_1}h_{\beta_1}\rangle+\langle\delta(\sigma_i,\alpha)\delta(\sigma_j,\theta)N_{\beta_2}h_{\beta_2}\rangle\nonumber
\end{equation}
and use the same procedure as in Eqs.~(\ref{n_vizinhos}-\ref{media_intermediaria}), one can put in terms of the probability of a given state $P(\alpha\beta\gamma\dots)$. For illustration, the third term of Eq.~(\ref{media_2}) will involve terms of the type:
\begin{equation}
\begin{split}
    &\left[\langle\delta(\sigma_i,\alpha)\delta(\sigma_j,\theta)N_{\beta_i}N_{\beta_j}N_{\beta_i}\rangle-\langle\delta(\sigma_i,\alpha)\delta(\sigma_j,\theta)N_{\beta_i}N_{\beta_j}\rangle\right]\frac{h_{\beta_1} h_{\beta_2}^2}{2}\\
    &+ \left[\langle\delta(\sigma_i,\alpha)\delta(\sigma_j,\theta)N_{\beta_i}N_{\beta_i}N_{\beta_i}\rangle-3\langle\delta(\sigma_i,\alpha)\delta(\sigma_j,\theta)N_{\beta_i}N_{\beta_i}\rangle-\langle\delta(\sigma_i,\alpha)\delta(\sigma_j,\theta)N_{\beta_i}\rangle\right]
\end{split}
\end{equation}
Second-order terms will leave results of the type:
\begin{equation}
\begin{split}
    &\langle\delta(\sigma_i,\alpha)\delta(\sigma_j,\theta)v_{\beta_i}v_{\beta_j}\rangle = \sum_{k, l}^{N_{\beta_i},N_{\beta_j}}\langle\delta(\sigma_i,\alpha)\delta(\sigma_j,\theta)\delta(\sigma_k,\beta_i)\delta(\sigma_l,\beta_j)\rangle=\\
    &=v_1(v_2-1)P_{i,j,k,l}(\alpha\theta\beta_1\beta_2) + v_iP_{i,j,k}(\alpha\theta\beta_i),
\end{split}
\end{equation}
Stating that $v_2\geq v_1$ to make the counting of neighbors true, and separating into the cases when $k\neq l$ (first term) and $k = l$ (last term).
Terms of third order or superior will have similar analysis, resulting in:
\begin{equation}
    \begin{split}
    &\langle\delta(\sigma_i,\alpha)\delta(\sigma_j,\theta)N_{\beta_i}N_{\beta_j}N_{\beta_i}\rangle = \sum_{k, l, m}^{v_{\beta_i},v_{\beta_j},v_{\beta_i}}\langle\delta(\sigma_i,\alpha)\delta(\sigma_j,\theta)\delta(\sigma_k,\beta_i)\delta(\sigma_l,\beta_j)\delta(\sigma_m,\beta_i)\rangle=\\
    &=v_1(v_1-1)(v_2-2)P_{i,j,k,l,m}(\alpha\theta\beta_1\beta_1\beta_2) + v_i(v_i-1)P_{i,j,k,l}(\alpha\theta\beta_i\beta_i)\\
    &+v_i(v_j-1)P_{i,j,k,m}(\alpha\theta\beta_i\beta_j)+v_iP_{i,j,k}(\alpha\theta\beta_i),
    \end{split}
\end{equation}
Higher-order terms require similar calculations. Summing up all the terms on Eq.~(\ref{media_2}), one can find the series:
\begin{equation}
\begin{split}
&\langle \delta(\sigma_i, X)\delta(\sigma_j,\theta)[1 - (1 - h_{\beta_1})^{N_{\beta_1}}(1 - h_{\beta_2})^{N_{\beta_2}}] \rangle =\\
&\left[ v_{\beta_1} h_{\beta_1} P_{i,j,k}(\alpha\theta\beta_1) - \frac{h_{\beta_1}^2}{2} v_{\beta_1}(v_{\beta_1} - 1) P_{i,j,k,l}(\alpha\theta\beta_1\beta_1) + \ldots \right] \\
+ &\left[ v_{\beta_2} h_{\beta_2} P_{i,j,k}(\alpha\theta\beta_2) - \frac{h_{\beta_2}^2}{2} v_{\beta_2}(v_{\beta_2} - 1) P_{i,j,k,l}(\alpha\theta\beta_2\beta_2) \ldots \right] \\
+ &\left[ -h_{\beta_1} h_{\beta_2} v_{\beta_1}(v_{\beta_2} - 1) P_{i,j,k,l}(\alpha\theta\beta_1\beta_2) + \frac{(h_{\beta_1})^2 h_{\beta_2}}{2} v_{\beta_1}(v_{\beta_1} - 1)(v_{\beta_2} - 2) P_{i,j,k,l,m}(\alpha\theta\beta_1\beta_1\beta_2) \right. \\
&\quad \left. + \frac{(h_{\beta_1})^2 h_{\beta_2}}{2} v_{\beta_1}(v_{\beta_2} - 1)(v_{\beta_2} - 2) P_{i,j,k,l,m}(\alpha\theta\beta_1\beta_2\beta_2) + \ldots \right].
\end{split}
\end{equation}
The next step is to choose one of the approximations Eq.~(\ref{approxim.}), and identify the respective series. In this case, the series will be:

\begin{equation}
   F(a,b) =  1-(1-b)^{v_{\beta_2}-v_{\beta_1}}(1-a-b)^{v_{\beta_1}-1}
    \label{2-term}
\end{equation}
with $a = h_{\beta_1}P(\beta_1)$ and $b = h_{\beta_2}P(\beta_2)$ in the mean-field case and $a = \frac{h_{\beta_1}P(\alpha\beta_1)}{P(\alpha)}$ and $b = \frac{h_{\beta_2}P(\alpha\beta_2)}{P(\alpha)}$ for the pair approximation, noting that in the mean-field case the exponent in the last term of Eq.~(\ref{2-term}) is only $v_{\beta_1}$. The second equation of Eq.~(\ref{pa_auxiliary_functions}) shows $F(a, b)$ for mean field and pair approximation for the cases $\langle u\rangle, \langle x\rangle, \langle y\rangle, \langle z\rangle$, while equation Eq.~(\ref{2-term}) correspond to pairwise average.

\subsection{$\langle\delta(\sigma_i,\alpha)\delta(\sigma_j,\theta)[1-(1-h_{\beta_1})^{N_{\beta_1}}(1-h_{\beta_2})^{N_{\beta_2}}]f(\mu, h_{\beta_{1,2}})\rangle$}

The calculation of this average hold the function $f(\mu, h_{\beta_{1,2}})$, Eq.~(\ref{mutation function}). For convenience, we will rewrite it as $\langle\delta(\sigma_i,\alpha)\delta(\sigma_j,\theta)\chi q_{\beta_n}\rangle = \chi_n\langle\delta(\sigma_i,\alpha)\delta(\sigma_j,\theta)q_{\beta_n}\rangle$ where $\chi$ and $q_m$ are defined as:
\begin{equation}
    \chi = \frac{T(\alpha \to \beta)}{p_1 + p_2} = \frac{1 - (1 - h_{\beta_1})^{N_{\beta_1}} (1 - h_{\beta_2})^{N_{\beta_2}}}{2 - (1 - h_{\beta_1})^{N_{\beta_1}} - (1 - h_{\beta_2})^{N_{\beta_2}}},
\end{equation}
\begin{equation}
    q_{\beta_n} = p_{\beta_n} \mu + p_{\beta_m} \left( 1 - \mu \right) \; ; m, n \in \{1, 2\}, \; m \neq n.
\end{equation}
$p_{\beta_m} = 1-(1-h_{\beta_m})^{N_{\beta_m}}$ is the probability of the predator $\beta$ with foraging area $m$ being successful in preying, and $q$ accounts for the probabilities of the occurrence of mutation (or not) in the process of foraging. Since the averages of $p_{\beta_m}$ were already made in previous steps, we can write in the general form that:
\begin{equation}
    \langle\delta(\sigma_i,\alpha)\delta(\sigma_j,\theta)T(\alpha\rightarrow\beta_m)\rangle = \chi_m(\mu G_{m'}+(1-\mu)G_{m})
\label{last_term}
\end{equation}
where $G_m$ refers to the average of $p$. Now, one must find $\chi_m$, but knowing that the joint probability of both $\beta_1$ and $\beta_2$ being successfully predated must be equal to the probability of $\alpha$ being predated at all, it is possible to write:
\begin{equation}
\begin{split}
        &\langle\delta(\sigma_i,\alpha)\delta(\sigma_j,\theta)T(\alpha\rightarrow\beta_1)\rangle+\langle\delta(\sigma_i,\alpha)\delta(\sigma_j,\theta)T(\alpha\rightarrow\beta_2)\rangle =\\
        &=\chi_1G_{\beta_1}+\chi_2G_{\beta_2} = \langle \delta(\sigma_i, X)\delta(\sigma_j,\theta)[1 - (1 - h_{\beta_1})^{N_{\beta_1}}(1 - h_{\beta_2})^{N_{\beta_2}}] \rangle
\end{split}
\end{equation}
The simplest solution to $\chi_1$ and $\chi_2$ is of the form:
\begin{equation}
    \chi_1 = \chi_2 = \bar{\chi} = \frac{1}{G_1+G_2}\langle \delta(\sigma_i, X)\delta(\sigma_j,\theta)[1 - (1 - h_{\beta_1})^{N_{\beta_1}}(1 - h_{\beta_2})^{N_{\beta_2}}] \rangle
\end{equation}

Thus, using the results derived earlier, its possible to find the final expression for Eq.~(\ref{last_term}), using Eq.~(\ref{2-term}) for pair-approximation:
\begin{equation}
    \langle\delta(\sigma_i,\alpha)\delta(\sigma_j, \theta) P(\alpha \to \beta_m) \rangle =  \frac{P(\alpha\theta)}{f_{\beta_1}+f_{\beta_2}}F\left(\frac{h_{\beta_1}P(\alpha\beta_1)}{P(\alpha)}, \frac{h_{\beta_2}P(\alpha\beta_2)}{P(\alpha)}\right) \left[\mu f_{\beta_{m'}} + \left( 1 - \mu \right) f_{\beta_m} \right],
    \label{3-term}
\end{equation}
with 
\begin{equation}
    f_{\beta_m} = 1-\left(1-\frac{\alpha\beta_m^t}{\alpha^t}\right)^{v_{\beta_m}-1}.
\end{equation}
Expression Eq.~(\ref{3-term}) is shown in the last term of Eq.~(\ref{pa_auxiliary_functions}). 
\bibliography{refs}

\end{document}